\documentstyle[12pt, a4]{article}
\begin{document}
\author{B. G\"{o}n\"{u}l and M. Ko\c{c}ak
\and Department of Engineering Physics, Faculty of Engineering,
\and University of Gaziantep, 27310 Gaziantep -T\"{u}rkiye}
\title{Remarks on exact solvability of quantum systems
with spatially varying effective mass}
\date{}
\maketitle

 Within the frame of a novel treatment we make a complete
 mathematical analysis of exactly solvable one-dimensional
 quantum systems with non-constant mass,
 involving their ordering ambiguities. This work extends
 the results recently reported in the literature and clarifies
 the relation between physically acceptable effective mass Hamiltonians.
\newline

\newline

The study of quantum mechanical systems with position dependent
mass has been raised some important conceptional questions, such
as the ordering ambiguity of the momentum and mass operators in
the kinetic energy term, the boundary conditions at abrupt
interfaces characterized by discontinuities in the mass function,
etc. Therefore, the form of the effective mass Hamiltonian has
been a controversial subject in the literature. In recent years
there has been a growing interest in the study of such systems due
to the applications in condensed matter physics and other areas
involving quantum many body problem. These applications stimulated
a lot of work in the literature regarding the development of
techniques for the treatment of such systems, for a recent review,
see [1-6] and the related references therein. In all these works
the main concern is in obtaining the energy spectra and/or wave
functions for quantum systems with spatially dependent effective
mass. Moreover, exact solvability requirements result in
constraints on the potential functions for the given mass
distribution. Though there has been a large consensus in favor of
BenDaniel and Duke Hamiltonian (BDD) \cite{BenDaniel} proposed in
the literature as an appropriate one, the question of the exact
form of the kinetic energy operator is still an open problem for
such systems.

Within this context, the present Letter involves an alternative
scheme to obtain unambiguously the Schr\"{o}dinger equation with
non-constant particle mass,  which makes clear the relationship
between the exact solvability of the Schr\"{o}dinger equation and
the ordering ambiguity. The model explored here restricts
naturally the possible choices of ordering and provide us a clear
comparison between the solutions of different but physically
plausible effective Hamiltonians clarifying  the physics behind
ambiguity.

To achieve our goal defined above, the recently developed
non-perturbative technique \cite{Ozer} is employed within the
frame of supersymmetric quantum mechanics \cite{Cooper}. In this
unified model, the BDD Hamiltonian is considered as an unperturbed
term while modifications due to other effective Hamiltonians are
treated as an additional potential in the same framework. This
realization is of prime significance in the calculation of
physical processes, which so far did not receive adequate
attention.

In the following sections the model used through the work is
introduced and its applications are presented where the
superiority of the present scheme is also discussed.

There are several ways to define the kinetic energy operator when
the mass is variable. Since the momentum and mass operators no
longer commute, the generalization of the Hamiltonian is not
trivial and this kind of physical problem is intrinsically
ambiguous. Starting with the von Roos effective mass kinetic
energy operator \cite{Von Roos}, which has the advantage of an
inbuilt Hermicity,

\begin{equation}
H_{\nu
R}=\frac{1}{4}[m^{\alpha}(\hat{z})\hat{p}m^{\beta}(\hat{z})\hat{p}m^{\gamma}(\hat{z})+
m^{\gamma}(\hat{z})\hat{p}m^{\beta}(\hat{z})\hat{p}m^{\alpha}(\hat{z})]+V(\hat{z}),
\end{equation}
where $\alpha+\beta+\gamma=-1$. By the correspondence in wave
mechanics $\hat{p}\rightarrow -i\hbar\frac{d}{dz},
~~\hat{z}\rightarrow z$ and on setting
\begin{equation}
m(z)=m_{0}M(z) ,~~~~~~   \hbar=2m_{0}=1 ,
\end{equation}
where $M(z)$ is the dimensionless form of the mass function, the
effective mass equation  can be written in a differential form,
\begin{equation}
-\frac{d}{dz}\left[\frac{1}{M(z)}\frac{d\Psi(z)}{dz}\right]+V^{eff}(z)\Psi(z)=E\Psi(z),
\end{equation}
Here, $V^{eff}(z)$ is termed the effective potential energy whose
algebraic form depends on the Hamiltonian employed
\begin{equation}
V^{eff}(z)=V_{0}(z)+U_{\alpha\gamma}(z)=V_{0}(z)-(\frac{\alpha+\gamma}{2})\frac{M''}{M^{2}}+
(\alpha\gamma+\alpha+\gamma)\frac{M'^{2}}{M^{3}}~~,
\end{equation}
in which the first and second derivatives of $M(z)$ with respect
to $z$ are denoted by $M'$ and $M''$, respectively. The effective
potential is the sum of the real potential profile $V_{0}(z)$ and
the modification $U_{\alpha\gamma}(z)$ emerged from the location
dependence of the effective mass. A different Hamiltonian leads to
a different modification term. Some of them are the ones of BDD
$(\alpha=\gamma=0)$, Bastard \cite{Gora} $(\alpha=-1)$,
Zhu-Kroemer (ZK) \cite{Zhu} $(\alpha=\gamma=-\frac{1}{2})$ and
Li-Kuhn \cite{Li} $(\beta=\gamma=-\frac{1}{2})$.

Considering the supersymmetric treatment of effective mass
Hamiltonians by Plastino and his co-workers \cite{Plastino}
\begin{equation}
A\Psi=\frac{1}{\sqrt{M}}\frac{d\Psi}{dz}+ W\Psi ,
~~~~A^{+}\Psi=-\frac{d}{dz}\left(\frac{\Psi}{\sqrt{M}}\right)+W\Psi
,
\end{equation}
where $A$ and $A^{+}$ are linear operators and $W(z)$ is a superpotential,
the supersymmetric Hamiltonians are expressed as
\begin{equation}
H_{1}=A^{+}A=-\frac{1}{M}\frac{d^{2}}{dz^{2}}-\left(\frac{1}{M}\right)'\frac{d}{dz}+W^{2}-\left(\frac{W}{\sqrt{M}}\right)',
\end{equation}
and
\begin{equation}
H_{2}=AA^{+}=H_{1}+\frac{2W'}{\sqrt{M}}-\left(\frac{1}{\sqrt{M}}\right)\left(\frac{1}{\sqrt{M}}\right)''.
\end{equation}
From which, supersymmetric partner potentials are
\begin{equation}
V_{1} ^{SUSY}=W^{2}-\left(\frac{W}{\sqrt{M}}\right)' ,~~~V_{2}
^{SUSY}=V_{1} ^{SUSY}+
\frac{2W'}{\sqrt{M}}-\left(\frac{1}{\sqrt{M}})(\frac{1}{\sqrt{M}}\right)''.
\end{equation}
At this stage, we use the spirit of recently developed
non-perturbative approach \cite{Ozer} by expressing the total wave
function as a product,
\begin{equation}
\Psi(z)=\Phi(z)\Theta(z).
\end{equation}
In the above equation, $\Phi$ denotes the wave function
corresponding to the unperturbed piece of the effective potential
in Eq. (4) while $\Theta$ is the moderating function due to the
modified term $U_{\alpha\gamma}$ therein.

The use of (9) in (3) yields
\begin{equation}
\frac{1}{M}\left(\frac{\Phi''}{\Phi}+\frac{\Theta''}{\Theta}+2\frac{\Phi'}{\Phi}\frac{\Theta'}{\Theta}\right)-
\frac{M'}{M^{2}}\left(\frac{\Phi'}{\Phi}+\frac{\Theta'}{\Theta}\right)=V_{eff}-E,
\end{equation}
which reduces to the usual Schr\"{o}dinger equation with a
constant mass when $M\rightarrow1$. With the consideration of (6),
where the superpotential now can be given as
\begin{equation}
W(z)=W_{0}(z)+\Delta W(z),
\end{equation}
with $W_{0}$ and $\Delta W$ being superpotentials corresponding to
the unperturbed potential $(V_{0})$ and modification term
$(U_{\alpha\gamma})$ respectively, Eq. (10) is transformed into a
couple of equation,
\begin{equation}
W_{0} ^{2}-\left(\frac{W_{0}}{\sqrt{M}}\right)'=V_{0}-E_{0} ,
~~~~W_{0}=-\frac{1}{\sqrt{M}}\frac{\Phi'}{\Phi},
\end{equation}
\begin{equation}
\Delta W^{2}-\left(\frac{\Delta W}{\sqrt{M}}\right)'+2W_{0}\Delta
W=U_{\alpha\gamma}-\Delta E, ~~~~\Delta
W=-\frac{1}{\sqrt{M}}\frac{\Theta'}{\Theta}.
\end{equation}
In the above equations, $E=E_{0}+\Delta E$ due to
$V_{eff}=V_{0}+U_{\alpha\gamma}$. Therefore one can easily see the
contributions, if any, to the energy and wave function due to the
use of effective Hamiltonians other than BDD which represents the
unperturbed Hamiltonian in the present scenario since it has no
modification term, see (4).

We are familiar with (12) as a standard supersymmetric treatment
of the Schr\"{o}dinger equation for the exact solutions. However,
Eq. (13) is new and is the most significant piece of the work
presented in this letter. Because it is a non-perturbative
approach by Riccati equation, which reproduces the whole
corrections  coming from $U_{\alpha\gamma}$ if, of course, Eq.
(13) is exactly solvable.

To proceed we remind a general consensus \cite {Alhaidari,Li} that
the resolution of the ordering ambiguity in this problem could
come from a scheme that starts with the relativistic Dirac
equation with spatially varying mass then taking the
non-relativistic limit. This is due to the fact that the Dirac
equation is inherently free from the ordering ambiguity and that
taking the non-relativistic limit is a well defined procedure.
Bearing in mind this point we propose a correct choice of $\Delta
W$ as
\begin{equation}
\Delta W=\left(\frac{\alpha+\gamma}{2}\right)\frac{M'}{M^{3/2}},
\end{equation}
which directs us to find correct ordering parameter(s) leading to
the physically plausible effective Hamiltonian(s). Through Eq.
(13), the parameters get decoupled in a natural way and the
ambiguity in the choice of proper kinetic energy operator
disappears. Substituting (14) into (13), we obtain
\begin{equation}
\Delta W^{2}-\left(\frac{\Delta
W}{\sqrt{M}}\right)'=U_{\alpha\gamma},~~~~ \Delta E=-2W_{0}\Delta
W,
\end{equation}
if either $\alpha=\gamma=0$ which yields the BDD Hamiltonian  or
$\alpha=\gamma=-\frac{1}{2}$ corresponding to the ZK Hamiltonian.
It is stressed that the results are independent of any choice of
$M(z)$ and in case $\alpha=\gamma=0$ Eq. (13) vanishes. This
restriction is in agreement with the discussion in Ref.
\cite{Morrow} and also with the work of Bagchi et all
\cite{Bagchi}.

Though the present formalism has a wide spread applicability, for
clarity we now simply consider the two examples which were
investigated in Ref.\cite{Plastino}. This consideration will shed
a light in understanding the interrelation between the BDD and ZK
effective Hamiltonians bearing in mind the results presented in
\cite{Plastino} for the systems of interest.

The simplest case of the shape invariance integrability condition
\cite{Cooper}, leading to exactly solvable potentials, corresponds
a uniform energy shift $\varepsilon$ between partner potentials,
\begin{equation}
V_{2} ^{SUSY}(z,\varepsilon)-V_{1}
^{SUSY}(z,\varepsilon)=\varepsilon=2E_{0}
\end{equation}
since $\Delta E$ term appearing in the partners due to
$U_{\alpha\gamma}$ cancels each other. The replacement of (8)into
(16) gives
\begin{equation}
\frac{2\left(W_{0}'+\Delta
W'\right)}{\sqrt{M}}-\left(\frac{1}{\sqrt{M}}\right)\left(\frac{1}{\sqrt{M}}\right)''=\varepsilon,
\end{equation}
from which one finds the superpotentials leading to the
hamiltonian with $V_{0}$,
\begin{equation}
W_{0}(z)=-\frac{1}{2}\left(\frac{1}{\sqrt{M}}\right)'+\frac{\varepsilon}{2}\int^{z}\sqrt{M(y)}dy,
\end{equation}
since $\Delta W=-(M'/2M^{3/2})=\left(1/\sqrt{M}\right)'$. To
finalize the full treatment, one needs the total superpotential,
$W=W_{0}+\Delta W$ from which the results in \cite{Plastino}, Eq.
(35) and the subsequent equations, can easily be reproduced.

From this short discussion, it is obvious that (i) there will be
no contribution to $E_{0}$ due to the modification term. For this
reason total energies in both system having a constant mass and
position dependent mass are equal. (ii) From (13), the
contribution of $U_{\alpha\gamma}$ to the unperturbed wave
function is (for the ground state)
\begin{equation}
\Theta_{n=0}(z)=exp\left(-\int^{z}\sqrt{M(y)}\Delta
W(y)dy\right)=m^{1/2}.
\end{equation}
Thus, going back to (9) along with Eqs. (12) and (18), the full
unnormalized ground state wave function is expressed as
\begin{equation}
\Psi_{n=0}(z)=\left[m^{-1/4}(z)\Phi(\bar{z})\right]m^{1/2}(z)=m^{1/4}(z)\Phi(\bar{z}),
\end{equation}
where $\bar{z}=\int^{z}\sqrt{M(y)}dy$, which supports the
reliability of the present formalism \cite{Gonul}. The excited
state wave functions can be determined \cite{Cooper} in algebraic
fashion by successive application of the linear operators in (5)
upon the ground state wave function. (iii) The both choice, namely
the BDD and ZK Hamiltonians are represented with a unique
superpotential leading to exactly equivalent wave functions for
shape invariant potentials. (iv) From (8), as
$\alpha=\gamma=-\frac{1}{2}$ , one gets
\begin{equation}
V_{2} ^{SUSY}=\left(V_{1}
^{SUSY}+U_{\alpha\gamma}\right)+\frac{2W'}{\sqrt{M}},
\end{equation}
pointing a duality between BDD and ZK schemes, which reveals the
suggestions in \cite{Gonul,Bagchi}.

Let us proceed with another example in Ref. \cite{Plastino} where
the superpotential leads to a Morse-like spectra,
\begin{equation}
W(z,A)=A+f(z),
\end{equation}
in which, within the frame of the present formalism,
$f(z)=f_{0}(z)+\Delta f(z)$ that turns the form of (22)into
\begin{equation}
W(z,A)=\left[A+f_{0}(z)\right]+\Delta f(z)=W_{0}+\Delta W(z)
\end{equation}
From the shape invariance condition $V_{2} ^{SUSY}(z,A)=V_{1}
^{SUSY}(z,A-\lambda)+R(A)$ used in the supersymmetric quantum
theory \cite{Cooper}, where $A$ is the potential parameter and $R$
involving both parameter, $A$ and $\lambda$ , leads to the ground
state energy of the system. In the light of the work carried out
in \cite{Plastino}, the substitution of (23) in (8) within the
frame of shape invariance condition above produces
\begin{equation}
\frac{2\left(f_{0}'+\Delta
f'\right)}{\sqrt{M}}-\frac{1}{\sqrt{M}}\left(\frac{1}{\sqrt{M}}\right)''=\lambda\left(\frac{1}{\sqrt{M}}\right)'-2\lambda\left(f_{0}+\Delta
f\right).
\end{equation}
Remembering $\Delta W=\Delta f=\left(\frac{1}{\sqrt{M}}\right)'$
for $\alpha=\gamma=-\frac{1}{2}$, the above equation is rearranged
as
\begin{equation}
f_{0}'(z)+b_{1}(z)f_{0}(z)=b_{2}(z),
\end{equation}
where
\begin{equation}
b_{1}=\lambda\sqrt{M},
~~b_{2}=-\left[\frac{\lambda}{2}\sqrt{M}\left(\frac{1}{\sqrt{M}}\right)'+\frac{1}{2}\left(\frac{1}{\sqrt{M}}\right)''\right].
\end{equation}
From (24) it is clear that $\Delta f$ term affects only $b_{2}$,
since when $\Delta f\rightarrow0~~b_{2}\rightarrow-b_{2}$. The
solution of differential equation in (25) gives
\begin{equation}
f_{0}(z)=\left\{C+\int^{z}b_{2}(y)dy~exp\left[\int^{x}b_{1}(t)dt~~\right]\right\}\times
exp\left[-\int^{z}b_{1}(y)dy~~\right],
\end{equation}
where $C$ is an integration constant. Employing the mass function
used in \cite{Plastino},
$M=\left[(\alpha+z^{2})/(1+z^{2})\right]^{2}$, we obtain
\begin{eqnarray}
W(z)=W_{0}+\Delta W= \nonumber
\end{eqnarray}
\begin{eqnarray}
=\left(A+Cexp\left[-\lambda
\left\{z+(\alpha-1)\arctan{x}\right\}\right]-\frac{z(\alpha-1)}{\left(\alpha+z^{2}\right)^{2}}\right)+2\frac{z(\alpha-1)}{\left(\alpha+z^{2}\right)^{2}},
\end{eqnarray}
that is Eq. (53) in \cite{Plastino}. From (12), the corresponding
potential function, energy and wave function can be expressed as
in \cite{Plastino}, which are out of interest in this letter.
Generalization of the above discussion to a formalism which is
applicable to all spatially varying masses, yields
\begin{equation}
W(z)=W_{0}+\Delta W=
\left\{A+Cexp\left[-\int^{z}b_{1}(y)dy~~\right]-\left(\frac{1}{2\sqrt{M}}\right)'\right\}+\left(\frac{1}{\sqrt{M}}\right)'
\end{equation}
Plastino and co-workers \cite{Plastino} studied this problem in
case $\alpha=\gamma=0$ considering only the BDD Hamiltonian and
arrived at Eq. (53) in their work, which addresses (28) in our
work. This means that BDD and ZK effective Hamiltonians in fact
reproduce same results employing an identical superpotential,
which once more supports the realization introduced by (21) that
they are their supersymmetric partners.

In this work we have discussed the problem of solvability and
ordering ambiguity in quantum mechanics for the systems with a
position dependent mass. The present scheme restricts the possible
choices of ordering. Proceeding with this consideration it has
been observed that the only physically allowable BDD and ZK
Hamiltonians are in fact their supersymmetric partners that
reproduce identical results in their independent considerations
due to use of an identical superpotential. We hope that this
observation would make a contribution to the ongoing debate in the
literature regarding the isospectral effective mass Hamiltonians.

 The authors wish to thank the referee
for his helpful comments and suggestions.

\end{document}